\documentclass[a4paper]{aa}

\def\gsim{\lower.4ex\hbox{$\;\buildrel >\over{\scriptstyle\sim}\;$}}
\def\lsim{\lower.4ex\hbox{$\;\buildrel <\over{\scriptstyle\sim}\;$}}
\renewcommand{\vec}[1]{\mbox{\boldmath $#1$}}
\def\Om{\Omega}
\def\q{\qquad}
\def\beg{\begin{eqnarray}}
\def\ende{\end{eqnarray}}
\def\gsim{\lower.4ex\hbox{$\;\buildrel >\over{\scriptstyle\sim}\;$}} 
\def\lsim{\lower.4ex\hbox{$\;\buildrel <\over{\scriptstyle\sim}\;$}}
\newcommand{\Ha}{\mbox{Ha}}

\renewcommand{\textrm} [1] {\rm #1} 
\renewcommand{\textit} [1] {\it #1}
\renewcommand{\textbf} [1] {\bf #1} 

\usepackage{graphicx}
\usepackage{txfonts}
 \begin{document}

   \title{The pinch-type instability of helical magnetic fields}
 \author{G.~R\"udiger\inst{1,2} \and M. Schultz\inst{1} \and D. Elstner\inst{1} 
          }

   \institute{Astrophysikalisches Institut Potsdam, An der Sternwarte 16,
              D-14482, Potsdam, Germany
            \and Forschungszentrum Dresden Rossendorf, P.O. Box 510119, D-01314 Dresden, Germany \\
              \email{gruediger@aip.de, mschultz@aip.de, delstner@aip.de}
      }

   \date{Received ; accepted }

  \abstract{}
 {To find out whether toroidal field can stably exist in galaxies    the current-driven instability 
 of toroidal  magnetic fields   is considered under the influence of an 
 axial magnetic field component  and under the influence of 
 both rigid and  differential rotation.}
 {The MHD equations are solved in a simplified model  with cylindric geometry. We assume  the  axial field  
 as uniform  and  the fluid  as incompressible.}
 {The  stability of a toroidal magnetic field is strongly influenced by   uniform  axial magnetic 
 fields.  If both field components are  of the same order of magnitude then the instability is  slightly supported 
 and modes with $m>1$  dominate. If the axial field even dominates   the most unstable modes have again $m>1$ but the 
  field is strongly stabilized.  All modes are suppressed by a fast rigid rotation  
 where   the   $m=1$ mode maximally resists. Just this mode becomes best re-animated 
 for $\Omega > \Omega_{\rm A}$ ($\Omega_{\rm A}$  the Alfv\'en frequency) if the rotation has a negative shear. 
  -- Strong indication has been found for a stabilization of the
 nonaxisymmetric modes for fluids with  small  magnetic Prandtl number if they are unstable for $\rm Pm=1$. 
 }
 {For  rotating fluids  the higher modes with $m>1$ do not play an important role in the linear theory. 
 In the light of our results galactic fields  should be marginally  unstable against perturbations with  $m\leq 1$. The corresponding growth rates are 
 of the order of the rotation period of the inner part of the  galaxy.}


 \authorrunning{G.~R\"udiger et al.}
 \titlerunning{Current-driven instability of rotating helical magnetic  fields}
\maketitle
\section{Introduction}
It seems to be an open question whether or not toroidal magnetic fields are stabilized  under 
the presence of poloidal field components.  Lundquist (1951)   formulated that  azimuthal 
fields with energy exceeding the energy of the axial field component become unstable. With other words, he found that {\em uniform 
axial fields act stabilizing} what -- if true -- would form a highly interesting finding  also for MHD experiments in the laboratory. 

By use of a cylindric magnetic geometry Roberts (1956) opened a new  discussion and found  for all ratios 
of the azimuthal field component and the axial field component instability against perturbations with high 
azimuthal mode numbers $m$. Tayler (1980)  re-discussed the adiabatic stability of stars with 
mixed poloidal and toroidal fields. For poloidal and toroidal field components of the same order 
he suggested  stability of the system but the final answer  to the question remained open. In his  detailed paper about magnetic instabilities Acheson (1978)  only considered  the stability of purely toroidal fields. Extending this work one can ask for the stability of a simplified magnetic  model where  axisymmetric  and stationary toroidal field are considered under the influence of a homogeneous axial field which itself is stable by definition. For an ideal medium Bonanno \& Urpin (2010) considered the stability of such  constellation without any rotation with respect to applications in  jet theory. They exclude stability for fields with pitch  $|B_z|/|B_\phi|$ of order unity. Particular attention is given to the instability of nonaxisymmetric modes with azimuthal mode numbers $m>1$. If the axial field dominates, the instability persists  for rather high  mode numbers (of order 100). The latter has  been found by use of an almost identical model by Tayler (1960) including the differences of the solution for different helicity of the background field.

A more complicated model has been investigated by  Braithwaite (2009) where also the poloidal field component (axisymmetric as also the toroidal one) can be unstable alone but the author finds stability of the combination of poloidal and toroidal field if both components are of the same order.

We shall show for dissipative fluids that compared to the case of purely toroidal fields the configuration with $|B_z|\simeq |B_\phi|$ is (slightly) more unstable while increasing $|B_z|$ more and more stabilizes the toroidal fields. With rotation the situation changes. If the rotation rate exceeds the Alfv\'en frequency of the toroidal field the instability of any $m$ is suppressed but the mode with $m=1$ persists at longest. If, as it is almost always the case, the rotation is not rigid then only the modes with small $m$ survive and the axisymmetric standard MRI starts to dominate for sufficiently fast rotation.  We find that the competition between the modes $m=0$ and $m=1$ should be observable with galaxies. Note that only a few examples of nonaxisymmetric magnetic field patterns have been found by the observers (Beck et al. 1996).

Probably, the hydromagnetic jets are  suitable subjects for the application of magnetic field instabilities  but for the low Reynolds numbers and Hartmann numbers which we shall deal with in the present paper the galaxies containing a remarkable  interstellar turbulence are forming better  objects.
Galaxies possess quadrupolar-type  magnetic fields with toroidal and poloidal components of the same 
order of magnitude ($\sim 10^{-5}$ Gauss) and with the phase relation $B_R B_\phi<0$. They rotate with the characteristic rotation law
\begin{equation}
  \Omega(R) = \frac{\rm const.}{R}
  \label{1}
\end{equation}
with $R\Omega\simeq \, 200$ km/s. Their characteristic density is of order $10^{-24}$ g/cm$^3$ and the magnetic diffusivity is about $10^{26}$ cm$^2$/s due to the action of interstellar turbulence. The resulting magnetic Reynolds number ${\rm Rm}\simeq U R_0/\eta$ is thus of order 1000 while the Lundquist number of the toroidal field ${\rm S}=B R_0/\sqrt{\mu_0\rho} \eta$ reaches values of 200. For galaxies both characteristic numbers are thus of the same order or more precisely  
\begin{equation}
{\Omega}\simeq 5\ {\Omega_{\rm A}},
\label{omaom}
\end{equation}
expressed with  the rotation rate and the Alfv\'en frequency $\Omega_{\rm A}$ (see Eq. (\ref{oma}), below). The question is whether such a magnetic constellation is stable against nonaxisymmetric disturbances with the azimuthal mode number $m$. 

The stability of a toroidal field strongly depends on its radial profile. So   the current-free profile $B_\phi \propto 1/R$ is stable against disturbances with the azimuthal mode numbers $m=0, 1, \dots$ . On the other hand, the profile $B_\phi\propto R$  is stable only against $m=0$ but it is unstable against disturbances with $m>0$ (Tayler 1957; Velikhov 1959). The latter profile will mainly be used in the present paper by its simplicity as it is due to a {\em homogeneous} axial electric current. A sufficiently strong toroidal field which is nearly uniform in radial direction is also unstable against disturbances with the mode number $m>0$. We shall model a galactic magnetic field  with respect to the equator by means of a Taylor-Couette flow with the rotation law (\ref{1})   periodic in the axial direction $z$. The axial electric current which produces the toroidal  magnetic field component is assumed as homogeneous in $z$. It is obvious that such a simple  cylindric model cannot describe the field geometry in a global rotating disk but only in one of its hemispheres. 

For simplicity the cylinders which confine the  conducting fluid are highly conducting, and no-slip  boundary conditions are used  at the cylinder walls. The magnetic background field (assumed as stationary) also possesses a uniform axial field component  so that the resulting field pattern forms a spiral. It is the stability of such a spiral with fixed current helicity which is considered in the present paper. With respect to galactic applications this is an
over-simplification as for dynamo-generated magnetic fields of either parity (with respect to the equator) the current helicity always behaves antisymmetric.

It appears to  be reasonable not to limit the azimuthal mode number to $|m|\lsim 1$ so that  also higher values can be considered (see Arlt et al. 2007; Bonanno \& Urpin 2010).

\section{The equations}
We are interested in the linear stability of the background field
$\vec{B}= (0, B_\phi(R), B_0)$, with $B_0=\rm const$, and the flow
$\vec{U}= (0,R\Omega(R), 0)$.
The perturbed  system is  described by
\beg
u_R, \ u_\phi, \ u_z, \ p, \ b_R, \ b_\phi, \ b_z.
\label{3}
\ende
Developing the disturbances into normal modes, the solutions
of the linearized MHD equations are considered in the form
\beg
f=f(R){\textrm{exp}}({\textrm{i}}(kz+m\phi+\omega t)),
\label{nmode}
\ende
where $f$ is any of the velocity, pressure, or magnetic field disturbances.

The resulting linear equations are
\begin{eqnarray}
\frac{\partial \vec{u}}{\partial t} + (\vec{U}\cdot\nabla)\vec{u}
 +  (\vec{u}\cdot\nabla)\vec{U}=
-\frac{1}{\rho} \nabla p + 
\nu \Delta \vec{u} + \\
+\frac{1}{\mu_0\rho}{\textrm{curl}}\ \vec{b} \times \vec{B}
+\frac{1}{\mu_0\rho}{\textrm{curl}}\ \vec{B} \times \vec{b},
\label{mhd}
\end{eqnarray}
\begin{eqnarray}
\frac{\partial \vec{b}}{\partial t}= {\textrm{curl}} (\vec{u} \times \vec{B})+  {\textrm{curl}} (\vec{U} \times \vec{b})+\eta \Delta\vec{b},
\label{mhd1}
\end{eqnarray}
and
\beg
{\textrm{div}}\ \vec{u} = {\textrm{div}}\ \vec{b} = 0,
\label{mhd2}
\ende
where $\vec{u}$ is the perturbed velocity, $\vec{b}$ the perturbed magnetic field,
$p$ the pressure perturbation and   $\nu$ and $\eta$ are the kinematic viscosity and 
the magnetic diffusivity. The magnetic Prandtl number is defined by
\beg
\rm Pm =\frac{\nu}{\eta}.
\label{pm}
\ende
The  stationary background solution is 
\begin{eqnarray}
\Omega=a_\Omega +\frac{b_\Omega}{R^2},  \ \ \ \ \ \ \ \ \ \ \ \ \ \ \ \ \ \ \
B_\phi=a_B R+\frac{b_B}{R},
\label{basic}
\end{eqnarray}
where $a_\Omega$, $b_\Omega$, $a_B$ and $b_B$ are constants defined by 
\begin{eqnarray}
\lefteqn{a_\Omega=\Omega_{\rm{in}}\frac{ \mu_\Omega-{\hat\eta}^2}{1-{\hat\eta}^2}, \quad \quad \quad \q
b_\Omega=\Omega_{\rm{in}} R_{\rm{in}}^2 \frac{1-\mu_\Omega}{1-{\hat\eta}^2},}
\nonumber\\
\lefteqn{a_B=\frac{B_{\rm{in}}}{R _{\rm{in}}}\frac{\hat \eta
( \mu_B - \hat \eta)}{1- \hat \eta^2},  \q\q \ 
b_B=B_{\rm{in}}R _{\rm{in}}\frac{1-\mu_B \hat\eta}
{1-\hat \eta^2}}
\label{ab}
\end{eqnarray}
with
\begin{equation}
\hat\eta=\frac{R_{\rm{in}}}{R_{\rm{out}}}, \; \; \; \quad\quad\q
\mu_\Omega=\frac{\Omega_{\rm{out}}}{\Omega_{\rm{in}}},  \; \; \;\quad\quad\q
\mu_B=\frac{B_{\rm{out}}}{B_{\rm{in}}}.
\label{mu}
\end{equation}
Here $R_{\rm{in}}$ and $R_{\rm{out}}$ are the radii of the inner and outer
cylinders, $\Omega_{\rm{in}}$ and $\Omega_{\rm{out}}$ are their rotation
rates, and $B_{\rm{in}}$ and $B_{\rm{out}}$ the azimuthal magnetic fields
at the inner and outer cylinders, resp.

The outer  value $B_{\rm out}$ is normalized with the uniform vertical field,
i.e.
\begin{equation}
\beta =\frac{B_{\rm out}}{B_0}.
\label{beta}
\end{equation}
As usual, the toroidal field amplitude is measured by the  Hartmann number
\beg
{\rm Ha} = \frac{B_{\rm out} R_0}{\sqrt{\mu_0 \rho \nu \eta}}.
\label{Ha}
\ende
Here $R_0=\sqrt{R_{\rm in}(R_{\rm out} - R_{\rm in})}$ is used as the unit of length,
$\eta/R_0$ as the unit of velocity and $B_{\rm in}$ as the unit of the azimuthal
fields. Frequencies, including the rotation $\Om$, are normalized with the inner rotation rate $\Om_{\rm in}$. The ordinary Reynolds number $\rm Re$ and the magnetic Reynolds number $\rm Rm$ are defined as 
\beg
{\rm Re}=\frac{\Om_{\rm in}  R_0^2}{\nu} \ \ \ \ \ \ \ \ \ \ \ \ \ \ \ \ \ \ \ \ {\rm Rm}=\frac{\Om_{\rm in}  R_0^2}{\eta},
\label{Rey}
\ende
and the  Lundquist number $\rm S$ is defined by 
$
\rm S=Ha \cdot \sqrt{\rm Pm} 
$ so that it can also be understood as the magnetic Reynolds number formed with the Alfv\'en frequency,
\beg
\Omega_{\rm A} = \frac{B_{\rm out}}{\sqrt{\mu_0\rho} R_0},
\label{oma}
\ende
instead of the rate of the global rotation. Note that the normalization concerns the maximal value of the toroidal field.

The boundary conditions associated with the perturbation equations are no-slip
for $\vec{u}$, i.e.
$
u_R=u_\phi=u_z=0,
$
and perfectly conducting for $\vec{b}$,
i.e. 
$
db_\phi/dR + b_\phi/R = b_R = 0.
$
These boundary conditions hold for both the inner and the outer cylinder.

All our calculations refer to a container with $R_{\rm out}=2 R_{\rm in}$, i.e. $\hat\eta=0.5$. For this choice a 
field which is current-free in the fluid is described by $\mu_B=0.5$. A homogeneous axial electric current between the cylinders requires $\mu_B=2$ which is the preferred value in this paper. The axial magnetic field component is assumed as uniform so that the resulting current helicity of the magnetic field is also homogeneous. This can only be  true within one hemisphere of the celestial body (here galaxy). Our 
model does not define an equator. We shall consider that hemisphere where the current helicity  $\vec{B}\cdot \vec{J}$ of the background field is negative, i.e.  
 $\beta<0$, which is  an arbitrary choice. As shown by R\"udiger et al. (2010) the resulting instability forms left  spirals if the rotation is slow. The kinetic helicity $\langle\vec{u}\cdot {\rm curl}\ \vec{u}\rangle$ of the perturbations (averaged over the azimuth) proves to be positive for this field. The kinetic helicity does not change its sign if the rotation is faster but then the  magnetic pattern forms right spirals.
\section{No rotation}
We start to consider a nonrotating container, i.e. $\rm Re=0$. In this case for given geometry 
and given  vector (0,$B_\phi(R), B_0$) of the magnetic field the critical Hartmann number does not depend on the magnetic Prandtl number $\rm Pm$. The   azimuthal drift of the nonaxisymmetric instability pattern vanishes (see R\"udiger \& Schultz 2010). Without rotation the instability patterns do not drift in azimuthal direction. For rapid rotation the drift rate always grows with $\Omega_{\rm in}$  rather than  with $B_\phi$ (see below).
\begin{figure}[htb]
   \centering
   \includegraphics[width=9.0cm]{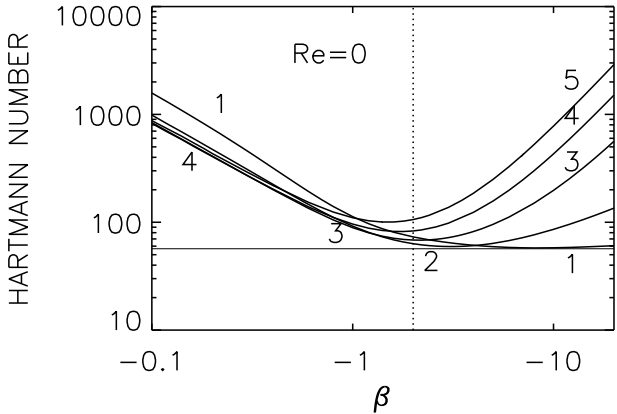}
   \caption{Neutral stability for negative $\beta$ and for $\mu_B=2$. The curves are marked with their mode number $m$. The minimum Hartmann number of the toroidal field with $\rm Ha=57.6$  exists for $|\beta|=8$ (thin horizontal line). The plot  valids for all magnetic Prandtl numbers.
              }
   \label{f1}
\end{figure}
\begin{figure}[htb]
   \centering
   \vbox{
    \includegraphics[width=9.0cm]{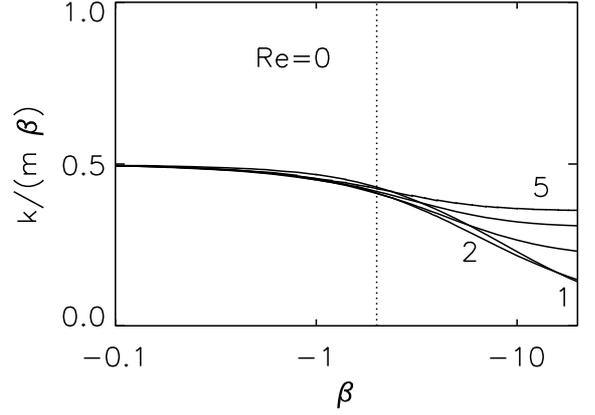}}
   \caption{The wave numbers normalized after (\ref{kk}) for  various  $\beta$. Note that the physical wave numbers run with the value of 
   $|\beta|$. $\mu_B=2$, $\rm Pm=1$. 
              }
   \label{f2}
\end{figure}

The most natural stationary  magnetic profile is $B_\phi \propto R$ which is the result of a homogeneous electric current flowing through the whole domain with $R<R_{\rm out}$ (see Roberts 1956). We know that for this case and for very large $|\beta|$ the critical Hartmann number has the value of 70.6  for $m=1$ (see R\"udiger et al. 2007).

This value of the critical Hartmann number is slightly reduced if a small and uniform  axial   component of the magnetic field is added to the system. Hence, a uniform axial field supports the pinch-type instability of the toroidal field. This effect, however, is rather weak: the critical Hartmann number sinks from about 70 to about 60 (see the  horizontal line  in Fig. \ref{f1}). For $m>1$ the destabilization  of the toroidal field by axial fields is much stronger so that for $|\beta|$  of order unity  all the modes with different $m$ have more or less the same  critical $\rm Ha$. We thus do not find a  stabilizing effect to toroidal fields by  axial fields components compared to fields of purely toroidal fields.

For $\beta=-8$ we find ${\rm Ha}=57.5$ as the absolute minimum of the stability curve for $m=1$. For stronger axial fields  the critical Hartmann number basically grows reaching values of about 1000 for $\beta \simeq -0.1$. For strong axial fields  the modes  $m>1$ possess lower critical Hartmann numbers than the mode with $m=1$. The differences of the curves with various $m$ are much smaller than those for weak $B_0$ but the Fourier component with $m=4$ possesses the lowest critical Hartmann number for $\beta=-0.1$. Nevertheless, for  dominating axial field the toroidal field is strongly {\em stabilized} -- in particular  the mode with  $m=1$. The modes with $m>1$ possess somewhat  slower critical Hartmann numbers but also these modes are basically stabilized (see  Fig. \ref{f1}). 

To summarize: The pinch-type instability of toroidal fields under the presence of a uniform axial magnetic field without rotation is strongly suppressed
by strong axial fields. The maximal stabilization happens for $m=1$, hence the  most unstable modes have azimuthal mode  numbers  $m>1$. 
If  $B_\phi$ and $B_z$ are of the same order then the field is  (slightly) more  unstable than for $B_z=0$. We find that with strong enough axial current-free  magnetic fields rather strong toroidal fields can be stored in the container. 

All results are invariant against the simultaneous transformation $m\to -m$ and $\beta \to -\beta$. 

The wave numbers $k$ of the unstable modes reflect the instability pattern. The shape of the cells is described by the relation
\begin{equation}
  \frac{\delta z}{R_{\rm out}-R_{\rm in}}=\frac{\pi}{k}.
  \label{k}
\end{equation}
The critical wave number for purely toroidal fields with $\mu_B=2$ is 2.8 so that after (\ref{k}) the cells are almost spherical. The wave numbers {\em with} axial field component  
are shown in Fig.~\ref{f2}. As expected they linearly grow for growing $|m|$ and for growing $|\beta|$. For dominating axial field the cells become longer and longer what is not unexpected.
For $\beta\lsim -1$ the simple relation
\begin{equation}
\frac{k}{m\beta}\simeq - 0.5 ,
\label{kk}
\end{equation}
results so that the aspect ratio of the cells  is more and more given by the pitch  of the field, i.e.
\begin{equation}
m \frac{\delta z}{\delta R} \simeq \frac{B_z}{B_\phi}.
\label{aratio}
\end{equation}
After Fig. \ref{f2} this relation is well-established for $|B_z|>|B_{\phi}|$.
\begin{figure}[htb]
   \centering
 \vbox{  \includegraphics[width=9.0cm]{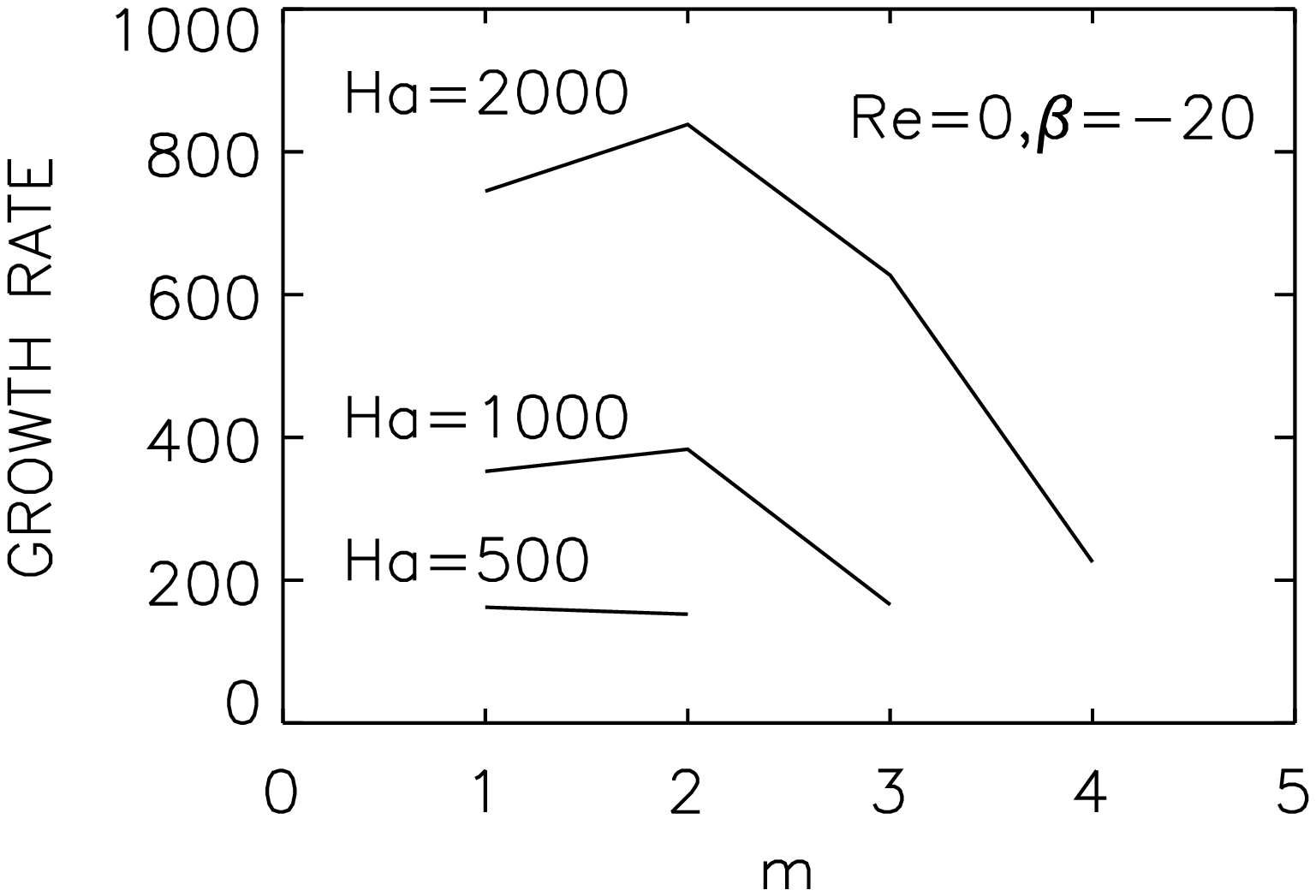}
   \includegraphics[width=9.0cm]{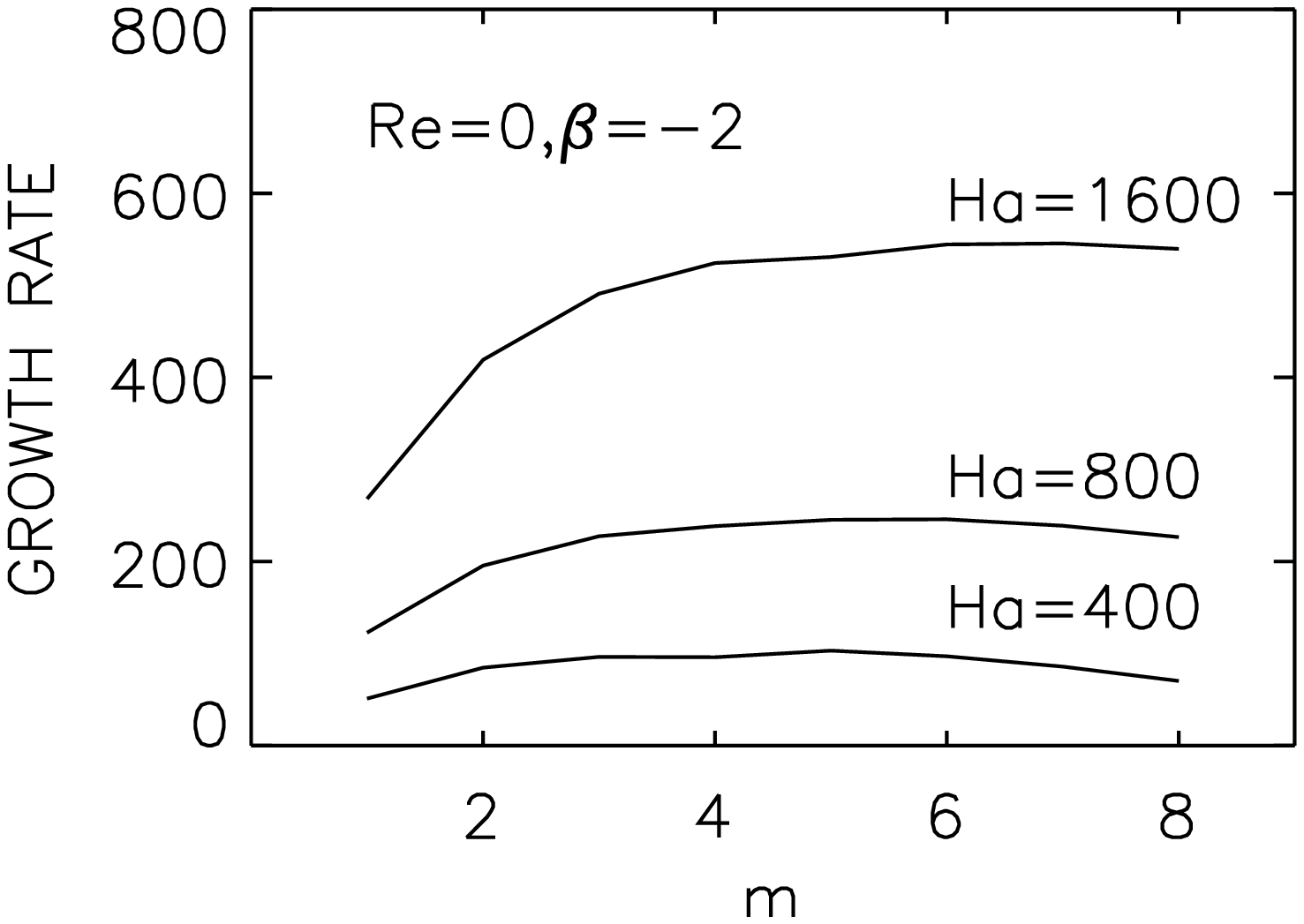}}
   \caption{No rotation: The growth rate in units of the diffusion frequency runs linearly with the Hartmann number of the toroidal field. Top:  $\beta=-20$. The mode with $m=2$  has the largest growth rate.  Bottom: 
   $\beta=-2$ (see the vertical lines in Figs. \ref{f1} and \ref{f2}).  The curves are marked with their mode number $m$. Generally, for finite $\beta$ maximum growth rates occur for   $m>1$.  $\mu_B=2$,  $\rm Pm=1$.         }
   \label{f3}
\end{figure}

The growth rates $\gamma=-\Im(\omega)$ must be given in units of the diffusion frequency $\eta/R^2_0$. We shall find that at least the modes with the higher $m$ are  strongly  dependent on the Reynolds number of rotation. Without rotation the growth rates for given $\beta$  and $\rm Pm$ are plotted in Fig. \ref{f3}. The used   $\beta$  close to the minimum where $|B_\phi| \simeq |B_z|$ will be the preferred value for many of the examples presented in this Paper.

Both plots in Fig.  \ref{f3} demonstrate the  finding that the growth rates run with the magnetic Alfv\'en frequency $\Omega_{\rm A}$. For stronger fields strong differences for the growth rates of various $m$ appear. For dominating azimuthal field ($|\beta|=20$, top) this is a weak effect but for $B_\phi$ and $B_z$ of the same order ($|\beta|=2$, bottom) it is strong. Of course, there are maxima; but for higher Hartmann number the highest growth rates belong to higher $m$.

The dependence of the growth rates on the magnetic Prandtl number is a complex problem.  The majority of the numerical simulations concerns to $\rm Pm=1$. In the Sect. 5.1 below we shall show that indeed this choice  forms a special case. For resting cylinders the  nonaxisymmetric mode with $m=1$ grows fastest for $\rm Pm=1$ if it is normalized with the geometrical average $\eta^*=\sqrt{\nu\eta}$ of  both  diffusivities. For given product of $\nu$ and $\eta$ the mode for almost equal diffusivities is most unstable while it becomes even stabilized if the two viscosities are too different.
The consequences for this finding may be dramatic if applied to numerical simulations. A field may be unstable for $\rm Pm=1$ which proves to be stable for more realistic very small or very large $\rm Pm$.

\section{Rigid rotation}
It is  known that rigid rotation stabilizes the magnetic perturbations.
This effect can easily be realized with our model. For the standard model with $\mu_B=2$ and ${\rm Pm}=1$ the growth rates have been calculated 
together with  the drift rates 
for a supercritical value of ${\rm Ha}$. 

\begin{figure}[htb]
   \centering
   \vbox{
   \includegraphics[width=9.0cm]{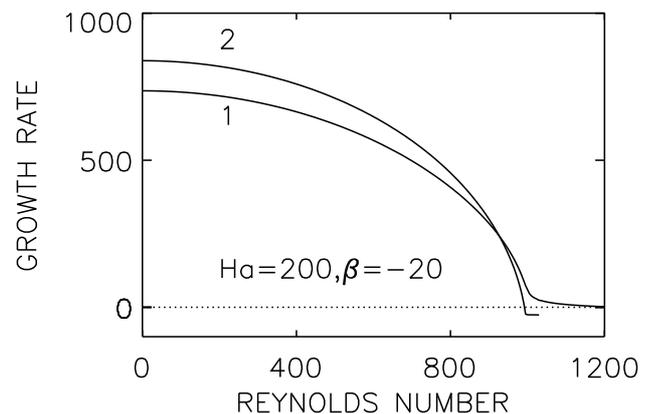}
}
   \caption{Rigid rotation and nearly toroidal fields ($\beta=-20$): The growth rates divided by the  diffusion frequency 
   for  supercritical magnetic field with  $\rm Ha=200$. The magnetic instability is stabilized for $\rm Re> Ha$. The kink-type mode with $m=1$ survives  rotation better than the higher modes.  $\mu_B=2$, $\beta=-20$, $\rm Pm=1$.
              }
   \label{f4}
\end{figure}

We start with a very small pitch, i.e. with nearly toroidal fields ($\beta=-20$). Figure \ref{f4} gives the results for  the growth rates normalized with the diffusion frequency. 
Both the given  modes for $m=1$ and $m=2$ are strongly suppressed by the basic rotation. We also find, however,  that the mode with $m=1$ better survives  the rotational suppression  than the modes with higher $m$.

\begin{figure}[htb]
   \centering
   \vbox{
   \includegraphics[width=9.0cm]{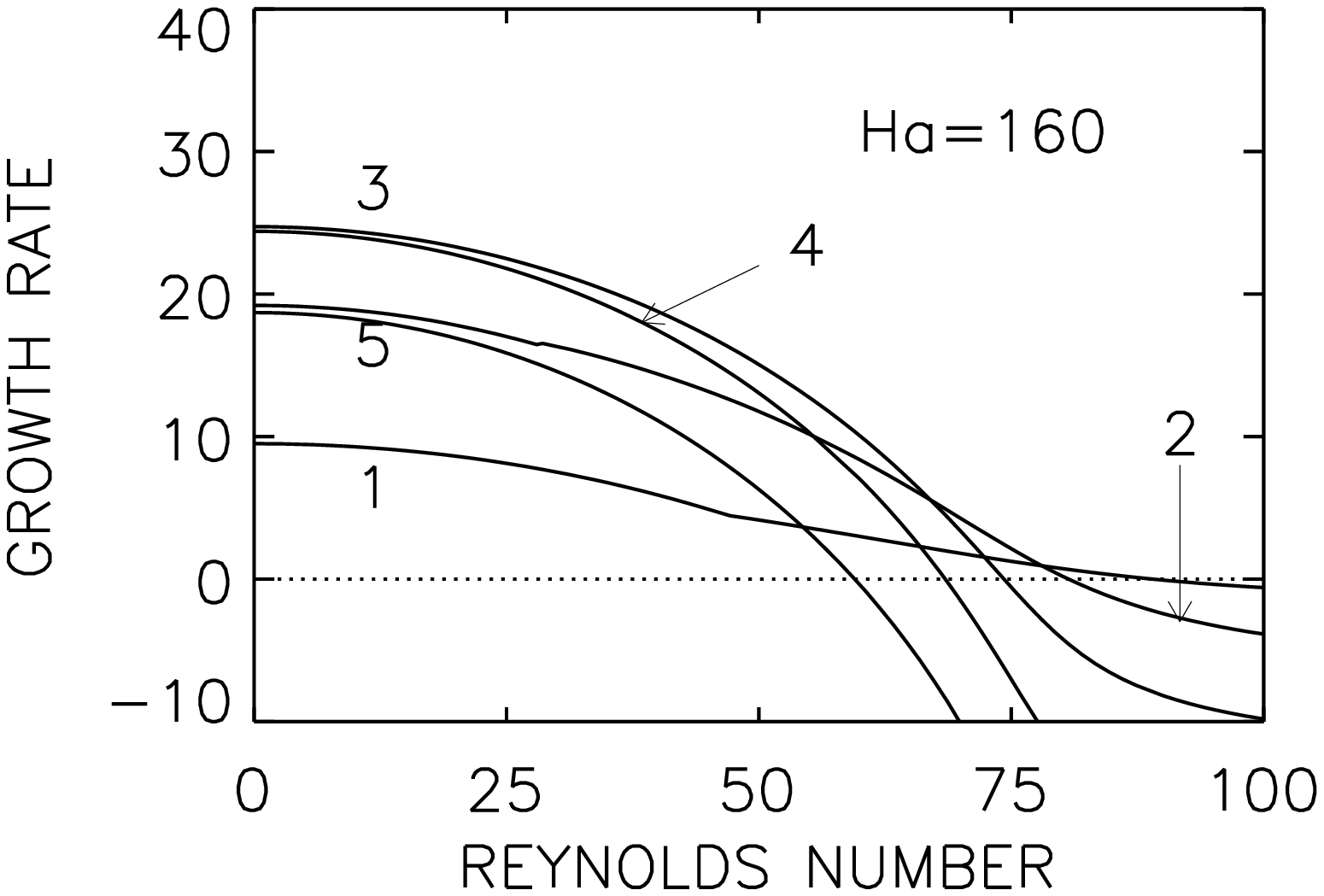}
    \includegraphics[width=9.0cm]{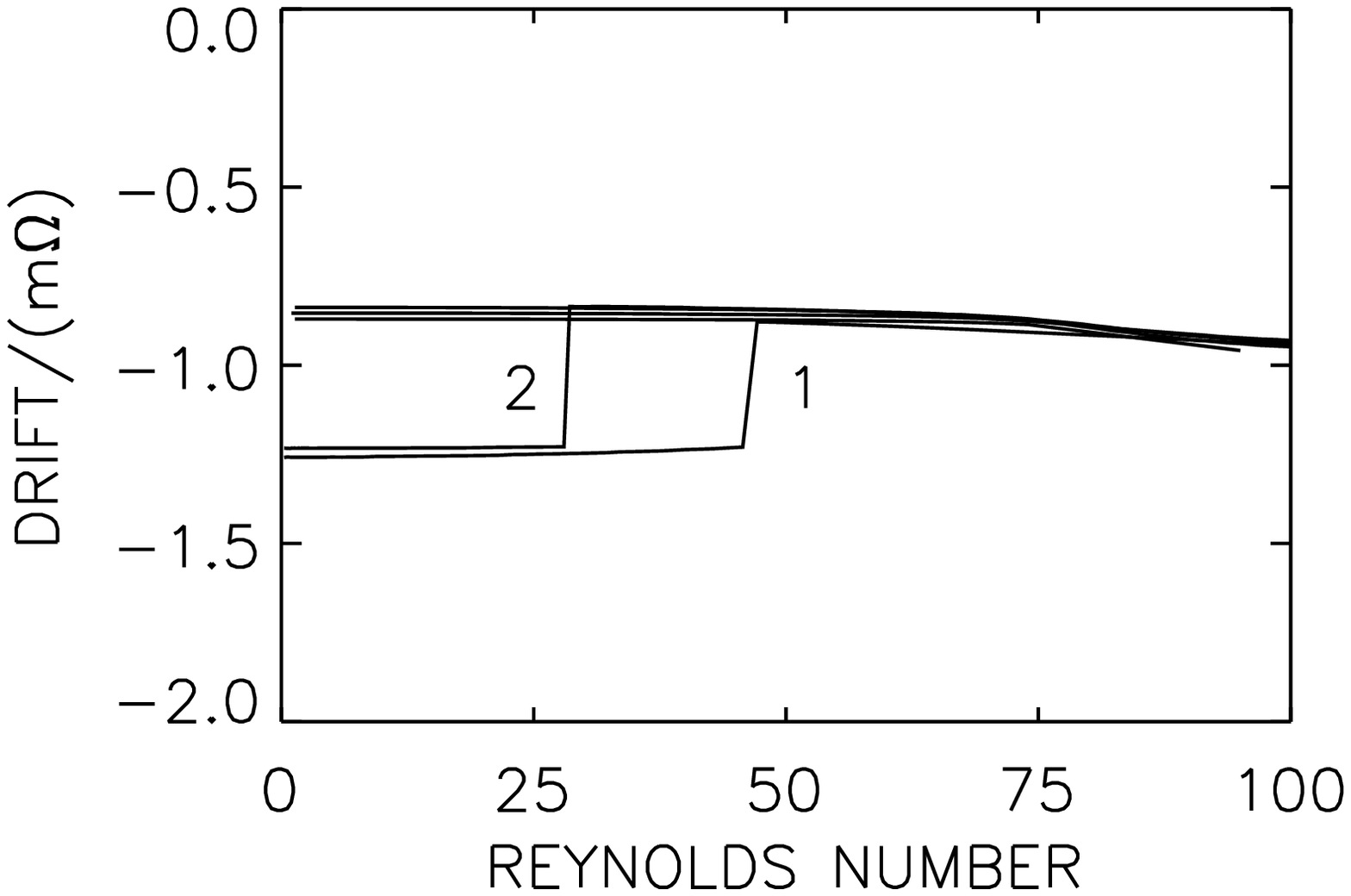}}
   \caption{Rigid rotation ($\mu_\Omega=1$): Top: Growth rates in unit of diffusion frequency. The magnetic instability is stabilized for $\Omega>\Omega_{\rm A} $. Bottom:
   the azimuthal drift after (\ref{drift}) of the modes.
     $\mu_B=2$, $\beta=-2$, $\rm Ha=160$, $\rm Pm=1$.}
   \label{f5}
\end{figure}

After Fig. \ref{f1} the most interesting  situation should exist  for magnetic fields with a pitch angle of order unity. The eigenvalues for the field with $\beta=-2$ have thus been calculated. Figure \ref{f5} gives the main results.
The critical ${\rm Ha}$ for $\beta=-2$ after Fig.~\ref{f1} is $\sim 70$.  One  finds positive growth rates  for slow rotation and stability for fast rotation. The instability cannot exist for ${\rm Re}>{\rm Ha}$ or in other words, for
\begin{equation}
\Omega>\Omega_{\rm A}.
\label{Oma}
\end{equation}
Again the mode with $m=1$ withstands at best the rotational suppression. It is also true that the modes with the highest  $m$ are suppressed already by lower  Reynolds numbers. Obviously,  {\em the dominance of the modes with  $m>1$ disappears  by  rigid rotation}.

For fast rotation the drift $\Re(\omega)/m\Omega$ of all modes approaches the value $-1$ so that after the relation
\begin{equation}
\frac{{\rm d}\phi}{{\rm d}t} = -\frac{\Re({\omega})}{m}
\label{drift}
\end{equation}
  an observer in the laboratory  system  finds the magnetic pattern as almost corotating. If the stellar rotation  can only  be observed via their magnetic variation then the rotation of such  an object is well approached by the rotation of the magnetic pattern. The observed (magnetic)  rotation is slightly slower than the real plasma   rotation. However, there is a jump in the curves: for slow rotation the modes with $m=1$ and $m=2$ rotate much faster than the container.
We have found such a  jump already for rotating stars with unstable toroidal fields (R\"udiger \& Kitchatinov 2010).

\section{Differential rotation}
There is a very new situation  if the outer cylinder rotates slower than the inner one.
 The simplified rotation law of our model may be the galactic one, (\ref{1}), so that for $\hat\eta=0.5$ the rotation ratio is $\mu_\Omega=0.5$ which also can be considered as the  normalized rotation of the outer surface of the container.

\begin{figure}[htb]
   \centering
    \includegraphics[width=9.0cm]{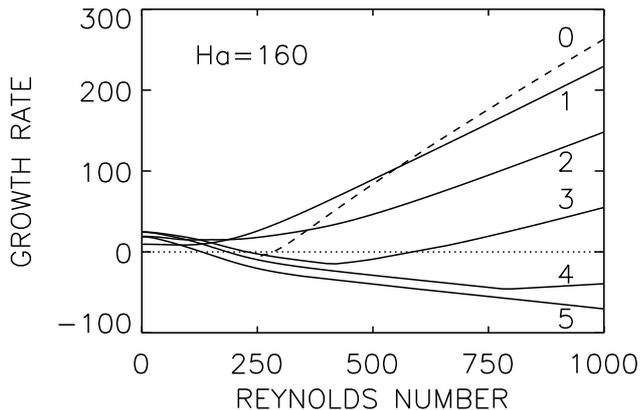}
   \caption{Differential rotation ($\mu_\Omega=0.5$): Growth rates in unit
    of diffusion frequency. The kink-type instability ($m=1$) is re-animated by fast rotation ($\Omega>\Omega_{\rm A} $).
     $\mu_B=2$, $\beta=-2$, $\rm Ha=160 $, $\rm Pm=1$.
              }
   \label{f6}
\end{figure}

The growth rates for ${\rm Ha}=160$ and the mentioned differential rotation are given in Fig.~\ref{f6} (top). The plot is identical with the plot for rigid rotation (Fig.~\ref{f5}, top) if the rotation is slow, i.e. for $\Omega\ll \Omega_{\rm A}$. All  modes are rotationally stabilized. For $\Omega \gsim \Omega_{\rm A}$, however,  the magnetic instability is re-animated but at most for the lower modes. Finally the $m=1$ mode becomes dominant; its growth rate (in diffusion units) becomes higher and higher finally running with the rotation frequency.  This new type of magnetic instability even exists for current-free toroidal magnetic fields so that we have named it  the azimuthal magnetorotational instability (AMRI). It is basically nonaxisymmetric with low $m$  and results from the  interaction of differential rotation and  toroidal fields (R\"udiger \& Schultz 2010). The growth rate $\gamma$ runs with $\Omega$ rather than $\Omega_{\rm A}$ when $\Omega>\Omega_{\rm A}$ (Fig.~\ref{f6}).

The higher modes  dominate only for small Reynolds numbers. They do not contribute to the instability for high Reynolds numbers as  they are damped by fast differential rotation. As we have shown in Sect. 3 the modes with $m>1$ are also damped for weak and for strong extra axial magnetic field components. Where they are most unstable (for $\beta$ of order unity) any rotation does suppress them. Their domain of dominance should thus be  small in astrophysical applications.
\subsection{Standard MRI}
Also axial fields can be  unstable under the presence of differential rotation. The leading mode of this standard MRI is axisymmetric (see Kitchatinov \& R\"udiger 2009). We have  given in Fig. \ref{f6}  also the growth rate of the mode $m=0$. It possesses the largest growth rate $\gamma$ if the rotation rate $\Omega$ is high enough, in this case the growth rate also runs with $\Omega$. We find that for $\Omega\gg\Omega_{\rm A}$  the most interesting case of $|B_\phi| \simeq |B_z|$ leads to a dominance of the standard MRI. Note the existence of an intersection between the growth rates of $m=0$ and $m=1$. Left from this point the nonaxisymmetric mode dominates the axisymmetric one while right of this point it is opposite. After Eq. (\ref{omaom}) galaxies do exist very close to that point. One should thus be aware that the stability of galactic fields should be rather delicate. We have shown that the negative shear of the rotation law strongly destabilizes the toroidal field. It is thus not clear, however, whether the most unstable mode is axisymmetric or not.

\subsection{The Pm-dependence of the growth rates}
The dependence of the growth rates on the magnetic Prandtl number $\rm Pm$ is not  trivial. We have  shown that for resting containers the characteristic Hartmann numbers for marginal instability do not depend on $\rm Pm$ (R\"udiger \& Schultz 2010). This is not true for the growth rates -- and this the more the faster the rotation is. Additionally, it is not obvious how to normalize the growth rates and the Reynolds numbers if both the diffusion times differ. We have also shown that the use of frequencies  normalized with a geometrically averaged  diffusion $\eta^*$ which is symmetrically formed with $\nu$  and $\eta$ with $\eta^*=\sqrt{\nu\eta}=\eta \sqrt{\rm Pm}$ seems to be most appropriate.
\begin{figure}[htb]
   \centering
   \vbox{
   \includegraphics[width=9.0cm]{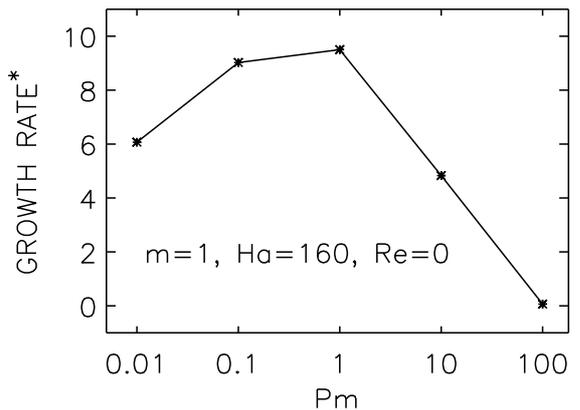}
    \includegraphics[width=9.0cm]{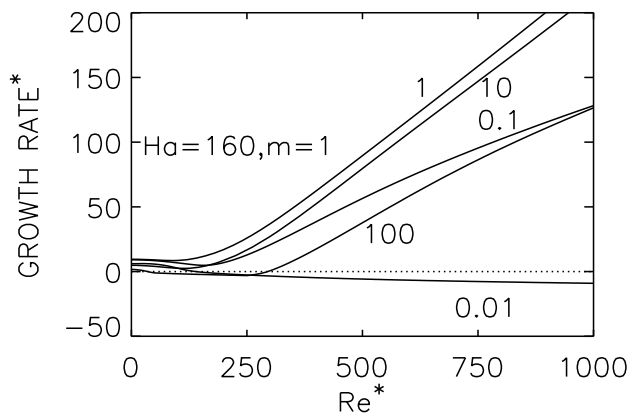}}
    \caption{The growth rates for the modes $m=1$ normalized with $\eta^*$ (see text) for fixed Hartmann number and  for various $\rm Pm$.
          Top: no rotation, Bottom: quasi-galactic  differential rotation ($\mu_\Omega=0.5$). The curves are marked with their magnetic Prandtl number.
  $\Ha=160$, $\beta=-2$, $\mu_B=2$. }
   \label{f7}
\end{figure}

In Fig. \ref{f7} the growth rates without and with (differential) rotation are given for a fixed Hartmann number. Both the growth rates of the mode $m=1$ and the global rotation rate are normalized with $\eta^*$, hence it is 
\begin{equation}
{\rm Rm^*}= \frac{\Omega_{\rm in} R_{\rm in}^2}{\eta^*}.
\label{Rem}
\end{equation}
One finds  $\rm Pm=1$ always leading to maximum growth rates for  slow and fast rotation. Either small or large magnetic Prandtl numbers lead to slower  growth of the instability than for $\rm Pm=1$. This effect is  so strong that the considered field pattern can even be {\em stabilized} if the magnetic Prandtl number is too 
small or too high. This is indeed the case in Fig. \ref{f7} for $\rm Pm< 0.01$. An instability  found with numerical simulations for $\rm Pm=1$ does not automatically exist for much smaller or much larger $\rm Pm$. If for a  given value of $\eta^*$ the magnetic field is unstable for $\rm Pm\simeq 1 $ this must not be true  if the numerical values of $\nu$ and $\eta$ are too
 different. This is an important restriction of the validity of  numerical simulations of magnetic stability/instability which are 
 operating with $\nu=\eta$. The  stability/instability of magnetic fields strongly depends on the magnetic Prandtl number of the fluid. For resting or rotating media   the fields are  most unstable for  $\rm Pm=1$.

Note that for given Hartmann number (here $\rm Ha=160$) one finds in Fig.  \ref{f7}  two regimes for the rotational influence on the growth rates. There is almost no influence of small $\rm Rm^*$ on the growth rate $\omega^*$. Figure   \ref{f7} (top) shows the {\em maximum} growth rate (for $\rm Pm=1$) as of order 10 leading to a {\em minimum} growth time of 0.1 diffusion times. For galaxies with $R_0\simeq$10 kpc and with $\eta\simeq 10^{26}$ cm$^2$/s the diffusion time is then 3 Gyr. One finds, however, fast global rotation accelerating the instability. From Fig.  \ref{f7} (bottom) the physical growth rate results as $\Omega/5$ so that the growth time is reduced by the rotation to about one rotation time. The rotation of the inner part of the galaxy is here concerned, with rotation times of about 50 Myr. Hence, the current-driven magnetic instability is a rather fast process.

\section{Almost homogeneous toroidal field}
\begin{figure}[htb]
   \centering
   \includegraphics[width=9.0cm]{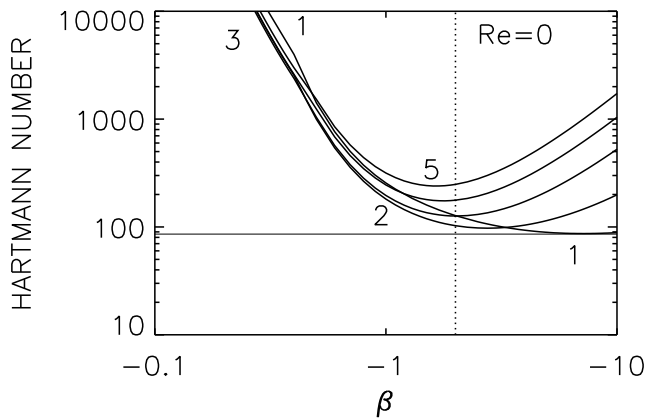}
   \caption{The same as in Fig. \ref{f1} but for $\mu_B=1$. The absolute minimum of the Hartmann number of $\rm Ha=86.8$ is located at $\beta=7.25$ for $m=1$. Note the extreme stabilization of $B_\phi$ for increasing $B_z$.
              }
   \label{f8}
\end{figure}

To check the consistency of our model in more detail the Figs. \ref{f1} and \ref{f6} are modified for  almost uniform toroidal fields, i.e. with $\mu_B=1$. We find very similar results but with slight numerical differences. Figure \ref{f8} reveals the stabilizing action of axial fields to the pinch-type instability of the toroidal field to be  much more effective for the case of almost homogeneous $B_\phi$. The critical Hartmann number for instability grows by orders of magnitudes if the axial field grows only by a factor of five. 

\begin{figure}[htb]
   \centering
    \includegraphics[width=9.0cm]{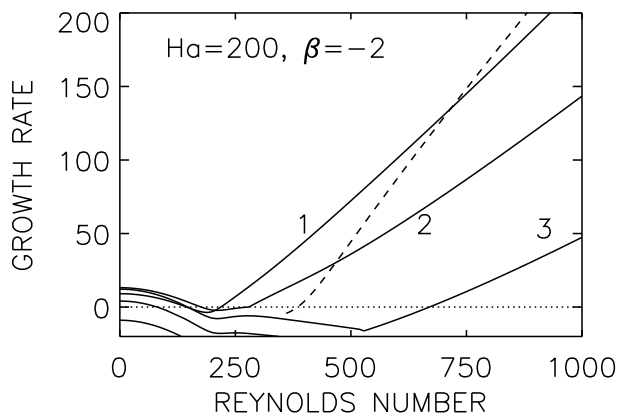}
   \caption{The same as in Fig. \ref{f6}  but for $\mu_B=1$. The dashed  line is for $m=0$ which  dominates for faster rotation. With the values characteristic for galaxies (right axis) one finds that  the axisymmetric standard MRI should dominate.
              }
   \label{f9}
\end{figure}

Also the complex influence of differential rotation on the instability of those fields with same order of toroidal and poloidal magnitude  shown by Fig. \ref{f6} exists for the case of almost homogeneous toroidal fields (Fig. \ref{f9}). Slow rotation acts stabilizing but  the modes with higher $m$  dominate for a while. For fast rotation the modes with low $m$ are re-animated as they form  the new instability. For slow rotation the modes with higher $m$ exhibit the maximum growth rates but for fast and differential rotation the mode with $m=1$ grows fastest.

Again the axisymmetric mode with $m=0$ is concerned which starts to dominate beyond the given crossing point. The coordinates of the intersection are nearly the same as  in Fig. \ref{f6}. 

Obviously, the above  findings about the influence of differential rotation on the stability of magnetic fields with spiral structure do not basically depend on the radial profile of the toroidal field.
It makes thus sense to call -- as we did -- this effect as the magnetorotational instability which originally only concerned current-free  toroidal fields ($\mu_B=0.5$, in our notation).

\section{Summary}
In a cylindric geometry  the pinch-type instability of axisymmetric and also magnetic spirals with finite current helicity $\vec{B}\cdot \vec{J}$ is considered under the influence of rotation. The field is formed by an unstable toroidal field  and a uniform axial field which is stable by definition.  The pitch of the  spiral is given by the inverse of $\beta=B_\phi/B_z$ which is negative for the considered lefthanded spirals. The larger the pitch  of the  background field the higher is the azimuthal Fourier $m$ of the mode with the largest growth rate.

The excitation of modes with low $m$ is here of particular interest. For small $|\beta|$ and without rotation  typically a mode with $m>1$  is excited with the largest growth rate. As stressed by Bonanno \& Urpin (2010) this phenomenon could have  consequences for the jet theory. As we have shown, however, a growing axial field stabilizes the toroidal field more and more. The critical Hartmann number grows by orders of magnitudes if $|\beta|$   reduces from order unity to order 0.1 (see Figs. \ref{f1} and \ref{f8}). Helical background  fields  with large axial field component are thus much more stable than purely toroidal fields  without finite $B_z$.

Also a global rotation stabilizes the pinch-type instability. Figure \ref{f5} shows for a magnetic field with almost equal field components ($\beta=-2$) how the rotation quickly stabilizes the modes with $m>1$ while the kink-instability ($m=1$) remains unstable for a little faster rotation. The growth rates of the modes are continuously reduced by growing Reynolds numbers. Generally, the helical background fields are stable against all nonaxisymmetric perturbations if $\Omega\gg\Omega_{\rm A}$. 

A very new situation results for nonrigid rotation. Figure \ref{f6} clearly demonstrates with a rotation law known from galaxies that for  $\Omega>\Omega_{\rm A}$ the growth rates after a characteristic minimum at  $\Omega\simeq\Omega_{\rm A}$ again reach positive and large values.  
A rotation law with positive shear will always stabilize the nonaxisymmetric instability. It has also been shown that under the presence of differential rotation with negative shear the toroidal field can become unstable even if there is no electric current in the container (see R\"udiger \& Schultz 2010). While for slow rotation  the modes with higher $m$ are most unstable it is for fast rotation the mode with $m=1$.

If the field possesses an axial component then under the influence of differential rotation with negative shear  the standard MRI  appears in form of an growing axisymmetric ($m=0$) roll. The lines of marginal instability for $m=0$ and $m=1$ are crossing so that for fast enough rotation the axisymmetric perturbation dominates. In any case we find that a spiralic magnetic field under the influence of differential rotation with negative shear appears to be extremely unstable (see Fig. \ref{f9} for a field nearly uniform in radius). As our model roughly reflects the magnetic geometry  in galaxies with SN-driven interstellar turbulence one should expect their dynamo-generated magnetic fields as rather  unstable. Nonaxisymmetric global field configuration might {\em not} be the exception.

Our model also allows the variation of the magnetic Prandtl number. Figure \ref{f7} shows a rather clear  situation. Its parameters only depend on the product of $nu$ and $\eta$, i.e. they are invariant against an exchange of $\nu$ and $\eta$. We find for fixed value of $\nu \cdot \eta$ strong differences of the lines for $\nu=\eta$ and $\nu\neq \eta$. The media with $\nu=\eta$  are more unstable than the media with $\nu\neq \eta$. Moreover, if the magnetic field is unstable for ${\rm Pm}=1$ it can even be stable for ${\rm Pm}\gg 1$ or ${\rm Pm}\ll 1$. The result requires care with the interpretation of numerical instability calculations if the considered medium has a magnetic Prandtl number  much smaller than unity. A magnetic field configuration which for a given Hartmann number and $\rm Pm=1$ results as unstable can be stable for much smaller or much larger $\rm Pm$.

\end{document}